# Enhancing Spectroscopy and Microscopy with Emerging Methods in Photon-Correlation and Quantum Illumination

Chieh Tsao[1,2], Haonan Ling[1,3], Alex Hinkle[1], Yifan Chen[1], Keshav Kumar Jha[1,2], Zhen-Li Yan[1], Hendrik Utzat[1,2]*.

**Affiliation(s):**
[1]*Department of Chemistry, University of California, Berkeley, Berkeley, CA, USA.*
[2]*Materials Science Division, Lawrence Berkeley National Laboratory, Berkeley, CA, USA.*
[3]*Department of Mechanical and Aerospace Engineering, University of Central Florida, Orlando, FL, USA.*
Corresponding author: *hutzat@berkeley.edu

**Abstract.** Quantum optics has driven major advances in our ability to generate and detect correlations between individual photons. Its principles are now increasingly translated into nanoscale characterization techniques, enhancing spectroscopy, microscopy, and metrology. In this Review, we highlight rapid progress in the field driven by advances in single-photon detectors and quantum light sources, including time-resolved single-photon counting cameras, superconducting nanowire detectors, and increasingly bright sources of entangled photons. We emphasize emerging applications in super-resolution microscopy, measurements below classical noise limits, and photon-number-resolved spectroscopy—a powerful paradigm for probing nanoscale electronic materials and molecular dynamics. We conclude by outlining key technological challenges and future opportunities across materials science and bio-nanophotonics.

## Introduction.

Optical characterization tools have been key drivers of science for centuries by unlocking information across time-, frequency-, and length-scales. The ultimacy of information is encoded on the level of individual photons, naturally rendering method development in the regime of quantum optics a frontier-endeavour.

Detecting the correlations in energy, time, and polarization between individual photons lies at the heart of these methods, which continues to benefit from the technological developments in single-photon detectors[1–7] and entangled photon sources[8–11] over the last decade. With innovations like superconducting nanowire single-photon detectors (SNSPDs), single-photon response times down to picoseconds, minimal noise down to below one dark photon count per second, and broadened spectral range are now at researcher's disposal. Measurement of classical and quantum correlations of photons are becoming possible from the UV to the mid-IR spectral range[12,13], for instance covering blue emission aromatic residues in biomolecules and the vibrational fingerprint region.

Particularly powerful is the trend of multi-dimensional detector arrays for single-photon counting that readily expand the dimensionality in photon-correlation from time, energy, and polarization to *space*. *Spatio*-temporal and higher-order correlations offer exciting possibilities, for example in quantum microscopy beyond the diffraction limit and nanomaterials characterization[14,15]. In parallel, advances in nonlinear optics, such as improved nonlinear materials and techniques for spontaneous parametric down-conversion (SPDC), have boosted the brightness and practicality of quantum light sources, particularly of entangled photon pairs (BOX 1)[8,16,17].

This convergence of detector and source advancements is rapidly enabling new photon-correlation methods that bridge quantum optics with microscopy and spectroscopy and increasingly moving beyond proof-of-principle toward real-world use. These methods fall into one of two broad categories: 1) resolving temporal, spatial, and frequency photon correlations from sample emission or scattering[18,19]; and 2) illuminating samples with quantum light of defined correlations[20]. Both approaches can demonstrate notable advantages in resolution and signal-to-noise and unlock otherwise hidden observables.

Photon-resolved and quantum light methods (BOX 1) in spectroscopy and microscopy sprout from earlier work in the field of quantum optics, which focused on, for instance, explaining the fundamentals of quantum mechanics[21], optical cavities[22], and entangled photon generation in bulk materials[8] or cold atoms[23]. This initially disciplinary scope expanded substantially by merging with other areas, such as condensed phase materials of increasing complexity—for example at the intersection of quantum optics with semiconductor quantum wells[24], and later quantum defects[25]. Continuing this evolution, quantum optical concepts are increasingly applied to even more structurally complex systems, including electronic nanomaterials[26] and even cells[27].

This Review highlights the emerging opportunities in spectroscopy and microscopy made possible by incorporating quantum optical principles. We argue that the promotion of quantum optics from science to technology has a positive backaction into the basic sciences proliferated by this "technology push". By outlining the foundations of quantum optics and related technologies, we aim to equip the reader to assess the utility and challenges of current and future methods within their own research.

## Emerging Methods Analyzing Sample Photon Emission Statistics

Building on the generic Hanbury Brown-Twiss (HBT) setup (BOX 1), which measures the normalized second-order photon-correlation, $g^{(2)}(\tau)$, current technique development expands the photon-correlation paradigm to higher-order correlations in time, frequency, and space. The resulting methods can provide access to new observables of materials properties, higher imaging quality, super-resolution, or multi-modality from combined electron-photon correlations in materials characterization.

### *Applications of second-order correlation measurements*

Simple second-order correlation measurements are widely used for characterizing single-photon emitters in various systems, including atoms[28], polymers[29,30], monolayers[31–35], and quantum dots (QDs)[36,37], and provide valuable insights into the excited states of electronic materials.

Various material properties can be extracted from the antibunching dip observed at $g^{(2)}(\tau = 0)$ (BOX 1), which suggests the absence of simultaneous two-photon detection. The depth of the dip

(or, visibility) quantifies the purity of the single-photon state of the emitter as a key performance metric in quantum light generation. Deviation from the ideal unity suppression of two-photon emission can be tied to the emitter photophysics, such as emission from excitonic complexes i.e., multi-excitons in semiconductor quantum dots. In such cases, the dip also quantifies the emission quantum efficiency[38–42]. Additionally, visibility can serve as a quantitative measure of the number of single-photon emitters within the excitation volume[43,44]. An intriguing scenario arises when a long-lived ground exciton state is present, which facilitates the generation of biexcitons. These biexcitons subsequently decay into bright-exciton-correlated photon pairs, resulting in photon bunching at $g^{(2)}(0)$[45]. Additionally, the slope can be used to estimate fluorescence lifetimes[46], and oscillations or bunching features around the dip (the 'shoulders') can encode information on Rabi oscillation frequencies,[44] or the presence of long-lived states during the photon cycle[47].

The $g^{(2)}(\tau)$ function is also utilized to study emitter aggregates exhibiting cooperative emission phenomena e.g., superradiance[48,49] or superfluorescence[50,51]. These effects produce a highly directional "burst" of light manifesting as bunching in the $g^{(2)}(0)$ (BOX 1). Like the antibunching scenario, the slope of the bunching peak provides information on the radiative decay time of the coupled states of the aggregates, which might help unveil the origins of cooperative emission in some materials. Bunching from cooperativity has been observed in a wide variety of systems, including molecular gases[52] and aggregates[53–56], photosynthetic pigments[57,58], epitaxial[59–61] and colloidal[62,63] quantum dots, lanthanide-doped upconverting nanoparticles[64], diamond vacancy centers[65], superlattices[66], and thin films[67,68]. Interestingly, bunching in the $g^{(2)}(0)$ sometimes is observed in single emitters, such as perovskite nanocrystals, at low temperatures[45]. This phenomenon is likely due to the exciton-shelving effect, which is induced by a long-lived ground exciton state. This effect facilitates the generation of biexcitons, which subsequently decay into photon pairs that are correlated with bright excitons.

Time-resolved $g^{(2)}(\tau)$ measurements under pulsed excitation allow the correlation analysis of photons based on their arrival time relative to the laser pulse. This can, for instance, distinguish the pathways of energy transfer in bi- and multi-chromophore systems. Mechanisms such as – homo-Förster resonance energy transfer (homo-FRET), singlet-singlet annihilation (SSA), and singlet-triplet annihilation (STA), are involved in the energy transfer processes at different timescales in these systems with each exhibiting unique bunching or antibunching features[47,69–71]. By temporally selecting the photons for obtaining correlation, the time evolution of each energy transfer mechanism can be probed (Fig. 1a). This has been achieved for systems such as ladder-type poly(para-phenylene) nanoparticles[69] and multichromophoric DNA-origami[72].

Beyond these $g^{(2)}(\tau)$ measurements, accessing higher-order photon events can extend the study of photophysics to more complex many-body interactions.

*Higher-order and higher-dimension photon-correlation methods*

With multiple single-pixel detectors of increasing detection efficiency, the photon-correlation paradigm has been increasingly extended to higher orders, i.e., $g^{(n)}$ with n > 2[73,74]. Using n single-pixel point detectors enables the delineation of higher-order multi-exciton dynamics. For example, second-, third-, and fourth-order correlation measurements on individual CdSe/CdS nanoplates (Fig. 1b)[75] have found that the scaling of $g^{(3)}(0,0)$ in relation to $g^{(2)}(0)$ significantly deviates from a well-accepted collision model for exciton-exciton annihilation. This discrepancy is attributed to previously overlooked many-body interactions in the nanoplates, which play a crucial role in multi-excitonic state relaxation.

Recently introduced single-photon avalanche diode (SPAD) array detectors (BOX 2)[3], which allow time-tagging with each pixel of the arrays, have further extended photon-correlation measurements from the temporal into the *spatial* and *spectral* domains, enabling higher-dimensional analyses. A technique known as spectroSPAD uses a diffractive grating in combination with a SPAD array for spectrally-resolved photon counting (Fig. 1c). The method performs photon-number resolved spectroscopy with spectral resolution. This approach has been adopted to investigate biexciton–exciton emission cascades and measure the biexciton binding energy in various systems, including CdSe/CdS/ZnS QDs[15], CdSe/CdS coupled QDs[76], and $CsPbI_3$ QDs[77] with clear delineation of state-specific emission spectra.

More broadly, adopting spectroSPAD in electronic materials characterization more broadly promises new insight into many-body interactions. However, SPAD array technology still displays substantial shortcomings, including crosstalk and high dark count rates (40 counts per second for good and thousands of counts per second for bad pixels), which impair time tagging performance and introduce artifacts if not properly corrected for[14]. Superconducting nanowire single photon detector (SNSPD) arrays-with superior detection efficiency and lower noise- offer a compelling alternative[78,79], but their cost remains prohibitive (BOX 2). As demands from the quantum photonics community continue to drive improvements in SPAD and SNSPD technology, we expect these tools to become more accessible. Once these barriers are reduced, high-dimensional correlation measurements could be more broadly applied to study photon-level nonlinearities in optical switching[80], multi-excitons in two-dimensional semiconductors[81], or strongly coupled excitonic systems[82].

*Photon-correlation enhanced optical microscopy for super-resolution*

Intensity correlation microscopy (ICM) leverages photon correlations to achieve spatial super-resolution in fluorescence imaging (Fig. 2a). Because photon correlation functions are inherently nonlinear–e.g., $g^{(2)}$ scales with the intensity squared – they allow the reconstruction of narrower point-spread functions (PSFs) and enhanced image resolution.

One established ICM technique is super-resolution optical fluctuation imaging (SOFI), which exploits temporal fluctuations of blinking fluorophores for achieving super-resolution in wide-field microscopy[83]. A single fluorophore with fluctuating emission shows positive intensity correlations (bunching) at time delays shorter than its characteristic fluctuation time (typically micro- to milliseconds; see Fig. 2a, "intensity statistics"). In contrast, multiple emitters with uncorrelated blinking exhibit reduced bunching. Measuring the amplitude of photon bunching pixel-wise thus enables localization of emitters with overlapping PSFs. Formally this is achieved by extracting the cumulant -- related to the nth-order correlation function $g^{(n)}(\tau)$ -- at each pixel and using it as the contrast for image reconstruction. Theoretically the PSF can be narrowed by a factor of $\sqrt{n}$, with a corresponding improvement in image resolution.

Several SOFI extensions have been introduced. Cross-correlation SOFI (XC-SOFI) improves resolution by calculating pixel-wise spatial cross-correlations and spatio-temporal cross-cumulants to produce an upsampled image with reduced effective pixel size[84]. Combining SOFI with image scanning microscopy (SOFISM)[85] adds the advantages of ISM, leading to resolution enhancement by a factor of $\sqrt{2n}$ relative to wide-field imaging. Further enhancement, up to four- to six-fold, can be achieved by integrating second-order SOFI with structured illumination, including sinusoidal and focused Gaussian pattern[86].

A more recent type of ICM is antibunching microscopy (AM). Similar to SOFI, AM uses spatio-temporal photon-correlation to narrow the PSF. However, it relies on photon-antibunching – the quantum signature of single-photon emission on nanosecond timescales—as image contrast. (Fig. 2a, "quantum statistics")[87,88]. Pixels detecting photons from a single fluorophore show maximal antibunching dips (theoretically 100% visibility), while pixels receiving photons from multiple fluorophores show reduced dips. Mapping the "missing" photon pairs across the image enables resolution enhancement by $\sqrt{n}$. When AM is combined with ISM in a quantum image scanning microscopy (Q-ISM) scheme, the total resolution is improved by $\sqrt{2n}$, as demonstrated through imaging of 3T3 cells labeled with QDs[89].

Another approach, stochastic frequency fluctuation super-resolution (SFSR) imaging[90], leverages uncorrelated spectral fluctuations –such as spectral diffusion-- to improve the resolution of single emitters by compiling spatially-resolved frequency-time photon correlations in the image plane (Fig. 2a, "spectral statistics"). Theoretically, this technique also enables resolution enhancement of $\sqrt{n}$, where n is the order of the correlation function.

While initial Q-ISM implementations relied on multiple single-pixel detectors[89], array-based event-resolved detectors are the natural fit for future intensity correlation and quantum microscopies[14]. These detectors capture full spatio-temporal photon statistics, enabling the simultaneous computation of multiple correlation-based contrast functions. In principle, this supports multidimensional quantum microscopy tailored to different use cases. For example, SOFI may be ideal for room-temperature, label-based super-resolution, whereas AM and SFSR may offer better contrast for quantum emitters with low intensity fluctuations at cryogenic temperatures.

*Multi-modal photon-correlation measurements in combination with other techniques*

Photon correlations can also be combined with non-optical methods, enabling multi-modal approaches. In this context, correlation measurements have been integrated with scanning electron microscopy (SEM)[91], scanning transmission electron microscopy (STEM)[92–94], and scanning tunneling microscopy (STM)[95], offering a broad set of observables for characterizing electronic nanomaterials.

In SEM and STEM, the interaction of the electron beam with the material can induce optical photon emission—a process known as cathodoluminescence (CL). The measured photon-correlation $g^{(2)}(\tau)$ (Fig. 2b) is determined by the interplay of (1) the multiplication factor of emitted photons per electron interacting with the sample[96], and (2) the properties of the emissive states. The transition from photon-bunching to antibunching reflects the density of quantum emitters implanted in a host matrix[97]. CL photon-correlation can also reveal excitation and emission efficiency, the lifetime of excited states, and average number of photons generated per electron[91,97]. When acquired in scanning mode with nanometer precision, two-dimensional $g^{(2)}(\tau)$ maps make CL-correlation a powerful supplement to electron beam imaging methods.

For example, GaN nanowires (NWs) of varying diameters have been studied via SEM-CL correlation imaging. Differences in CL intensity were attributed to variations in emission efficiency across NWs, while sub-structure intensity differences within a single NW were linked to geometry-induced variations in electron excitation efficiency[91]. In STEM, a direct--albeit stochastic—correlation between the transmitted electron and the generated photon can be established[94]. Unlike conventional CL, this method does not rely on photon bunching to extract emission lifetimes; rather, it uses the intrinsic timing correlation between the electron and photon. Demonstrated in nanodiamonds, this approach enables the extraction of lifetimes for both coherent and incoherent photon generation channels, offering mechanistic insight into the emission processes[94].

STM provides access to photon correlation measurements with atomic-scale resolution, using tip-induced luminescence. This technique enables the study of exciton, charge, molecular, and atomic dynamics with picosecond timescale and picometer spatial resolution[95], and opens the door to investigations of collective phenomena in-situ assembled atomic aggregates with atomic precision.

These examples showcase the untapped potential of higher-order and spectrally resolved photon-correlation in non-optical nanoscopy[98].

Thus far, we have focused on analyzing photon statistics emitted from a sample. In the following section, we turn our attention to the use of quantum light -- light engineered with specific photon correlations -- for sample illumination.

# Quantum light illumination in spectroscopy and microscopy

Illumination with quantum light of defined photon correlations enables super-resolution, ultra-low excitation power imaging, changed instrument form factor and increased robustness, as well as an improved signal-to-noise ratio (SNR). In the following subsections, we will explore different quantum states of light and their potential use cases. Entangled pairs form the basis for preparing bespoke the needed quantum states such as heralded single photons and N00N states (BOX 1).

## *Time-frequency entangled photon pairs*

### *Heralded emission spectroscopy*

Entangled photon pairs with defined temporal correlations can, for example, be used to measure fluorescence lifetimes without pulsed excitation lasers. The first of the two photons is directly detected and time-tagged and the second photon is interfaced with a sample. If the second photon evokes subsequent sample fluorescence, it is heralded by the detection of the first. The $g^{(2)}(\tau)$ between the heralding photon and the fluorescent (heralded) photon can effectively measure the time-delay between photon absorption and emission (Fig. 3a)[99,100]. The lifetime can then be obtained from the slope of the $g^{(2)}(\tau)$. This technique has only recently been demonstrated using organic dyes[99,100] and natural photosynthetic complexes[101]. Its simplicity – eliminating the need for pulsed or modulated laser source – may facilitate seamless integration with on-chip analytic devices of minimal form factor, for example useful in lab-on-a-chip applications or remote analytics in space exploration.

### *Entangled two-photon microscopy*

Heralded emission spectroscopy uses single-photon absorption. The temporal correlation in entangled pairs can also be used for two-photon excitation used in e.g., entangled two-photon (ETP) microscopy. ETP microscopy takes advantage of the temporal correlation between entangled pairs to lower the required illumination intensity by many orders of magnitude[102]. Compared to conventional two-photon microscopy, in which the probability of two-photon excitation scales quadratically with the excitation power, the signal intensity in ETP microscopy is linearly dependent on the intensity (flux) of entangled pairs. A possible quantum enhancement effect on the improved two-photon absorption cross-section in ETP was previously debated. However, recent work seems to point to a purely statistical explanation of the comparatively high overall excitation rates in ETP[103]. Fig. 3a compares the signal from classical and ETP microscopies indicating a six order of magnitude reduction in the required photon flux[104], which minimizes photobleaching and sample damage, rendering the ETP approach promising for sensitive biological and materials imaging, or applications requiring suppression of uncorrelated external background photons e.g. from experiment-specific secondary illumination or bioluminescence.

### *Hong-Ou-Mandel interference in microscopy and spectroscopy*

Two-photon pairs are also advantageous in phase microscopy and measurements. The corresponding methods rely on Hong-Ou-Mandel interference (BOX 1), which measures a dip in the $g^{(2)}(\tau)$ recorded at the output of a beamsplitter due to two-photon quantum interference. Analyzing the position, shape, and visibility of the dip measures the degree of two-photon self-similarity.

One prominent advantage of such HOM interference between two-photon pairs generated by narrow bandwidth pumping is its inherent dispersion cancellation that allows accurately assessing the phase shift of photons traveling through samples showing weak dispersion[105–107]. The preserved temporal correlation of the two-photon wavefunction effectively cancels out any dispersion-induced phase shifts, for example if one of the two photons passes through a weakly dispersive medium accumulating an extra phase relative to the second photon before interfering at the beamsplitter. As a result, the shape of the HOM dip remains unchanged while the position of the dip shifts in relation to the sample thickness[106,108]. This unique feature renders HOM interferometry a robust and precise tool for use cases where maintaining phase stability proves challenging, for example in biological samples exhibiting thermal fluctuations.

HOM interference can also be used in sensitive phase microscopy. By applying Fisher information analysis and maximum-likelihood estimation procedures to accurately locate the HOM dip[109], this approach has been used in low photon flux and scan-free microscopy of transparent samples with micrometer-level precision in depth profiles (Fig. 3b)[110]. Such advancements facilitate label-free imaging of light-sensitive materials or biological samples. Thanks to its high dynamic range, HOM microscopy is well-suited for samples with large spatial variations in phase difference.

Changes in HOM dip shape and anisotropy contain critical sample information e.g., the $T_2$ relaxation time of resonant optical transitions (Fig. 3c)[111]. The effect of HOM dip distortion under resonant entangled photon light-matter interaction has so far been shown in Nd:YAG crystals[112], Si nanodiscs on quartz[112], and IR-140 dye solutions[113], suggesting that multimodal HOM-based absorption/emission/phase contrast microscopy is within reach.

*Photon coincidence microscopy*

Photon coincidence microscopy (PCM) can achieve enhanced SNR and contrast compared to traditional wide-field or confocal scanning microscopy by discriminating stray- and background photons via temporal coincidence counting of entangled photons under low-flux illumination[114,115]. Ghost imaging (GI) and heralding imaging (HI), relying on distinct physical mechanisms, are two common PCM configurations (Fig. 4a). In GI, the signal (or idler) photons act as the heralding arm, passing through the object and being detected by a bucket detector (large area point detector). The idler (or signal) photons then illuminate a multipixel detector to form images using photon-correlation without traversing the sample. In HI, the object is placed in the heralded arm, and images are formed by the coincidence of photons registered in the heralding arm with those passing through the sample in the heralded arm. Compared to the direct imaging (DI) scenario, both GI- and HI-based PCM can offer significantly improved contrast, defined as the difference between the maximum and minimum intensities divided by the sum of the maximum and minimum intensities (~0.7 for both GI and HI while ~0.2 for DI). This advantage allows sample imaging with low photon fluxes, as validated by imaging of a wasp wing with 0.45 photons per pixel using a GI configuration[116]. Moreover, sub-shot-noise quantum imaging (SSNQI) can be achieved using a HI-based PCM by subtracting the "locally" correlated noise pattern[117]. Using nondegenerate entangled photon pairs in PCM may offer additional benefits, such as enabling imaging in hard-to-access wavelength regions to avoid the regions with low detector sensitivity (>1100 nm) or potential sample damage by (i) detecting the twin photon at different energies[118] and (ii) minimizing photodamage using lower-energy photons to illuminate the samples[119].

It is noteworthy that (quasi-)thermal light, which shows thermal distribution in photon statistics and bunding at $g^{(2)}(0)$, can also serve as illumination in such microscopy[120–122]; however, thermal light, being an incoherent statistical mixture of photon pairs, makes thermal ghost imaging more vulnerable to stray light. In contrast, entangled photon pairs form a pure state, allows photons to be distinguished from stray light, thereby typically providing superior SNR[123,124].

*Imaging with undetected photons*

Imaging with undetected photons (IUP)[125] shares some advantages with PCM, including the ability to image in the few-photon flux regime and in otherwise challenging spectral regions. IUP directly forms images via a distinct mechanism without coincidence counting. As shown in Fig. 4a, IUP uses two nonlinear crystals (NL1 and NL2) to generate photon pairs. The idler photon from NL1 interacts with the object, while the idler from NL2 does not. These two idlers overlap, creating interference that is modulated by the object. Since signal photons from both crystals are entangled with their corresponding idler photons, the interference caused by the idlers transfers information about the object to the signal photons. The signal photons, which do not interact with the object, can then reveal the object's image, even though the idler photons that interact with the object are never detected. This imaging without photon detection can be leveraged to capture mid-infrared (mid-IR) images, which are typically challenging to achieve due to limited mid-IR detection technology. A notable example includes wide-field imaging of a mouse heart using nondegenerate photon pairs consisting of signal photons in near-IR and idler photons in mid-IR range (3.4 to 4.3 μm)[126].

### *N00N states*

N00N state (BOX 1) illumination enables imaging with high SNR in different types of phase microscopies and offers a potential approach to beating the diffraction limit. A key characteristic of the N00N state is its oscillation faster than a coherent state, which shortens the period of interferometric fringes by a factor of N, the number of photons in the same mode. Using N00N states therefore cuts the effective wavelength of illumination by a factor of N and surpassing the diffraction limit[127–129]. Under ideal conditions, the interference fringes with shortened period may further lead to an N times sharper intensity jump across the edge of the fringes and thus enable the beating of the shot-noise limit, known as supersensitivity[129]. This supersensitivity finds applications in differential interference contrast (DIC) microscopy for examining opaque materials. Combining a laser confocal microscopic setup with DIC microscopy method, the SNR approximately 1.35 times better than the standard quantum limit using N = 2 N00N state as illumination (Fig. 4b)[130]. In addition, utilizing polarization-dependent N00N states with N = 2 and N = 3 to illuminate a quartz crystal fragment can lead to an $\sqrt{N}$-times enhancement in sensitivity compared to classical phase microscopy[131]. These findings underscore the potential of N00N states in facilitating highly sensitive microscopy. Moreover, the potential for superresolution motivates N00N state illumination in high-resolution photo-lithography, potentially expanding the scope of quantum light applications in nanoscience[132,133].

## *Hyperentangled photon pairs*

Hyperentangled photon pairs extend the utility of entangled pairs by introducing simultaneous entanglement in multiple degrees of freedom e.g., energy and polarization[134]. Imaging with hyperentangled pairs is emerging as a promising multi-modal technique, offering access to rich, multidimensional information[134,135]. For instance, simultaneous entanglement in the spatial mode, polarization, and energy has already enabled quantitative quantum birefringence imaging, allowing concurrent measurement of phase retardation and the principal refractive index axis. Using the HI configuration, a sub-shot-noise SNR was also achieved (Fig. 4c)[136]. Hyperentangled photon pairs have also been applied to large field of view and scan-free holography[137,138]. In one demonstration, phase images were encoded in spatial-polarization hyperentanglement and subsequently retrieved, enhancing spatial resolution by a factor of 1.84 compared to classical holography[138].

The expanded degrees of freedom in hyperentangled photon pairs may ultimately support quantum-limited multimodal imaging, combining microscopy and entangled-photon absorption measurements, and holography in a single measurement.

## *Squeezed light*

Squeezed light (BOX 1), a quantum state of light famously used in gravitational wave detection by LIGO[139], has the potential to surpass the shot-noise limit and thereby enhance the SNR in microscopy. For example, in stimulated Raman microscopy, two light sources—a pump for excitation and a squeezed light at the Stokes frequency for stimulation—have been used to image yeast cells in aqueous solution (Fig. 4d)[140]. The measured noise is reduced by 1.3 dB below shot noise, corresponding to a remarkable 35% improvement in SNR. By surpassing the shot-noise limit, this work stimulates exploration of the squeezed light technique to a range of applications, including video-rate imaging of weak molecular vibrations and label-free, spectrally resolved imaging.

Advances in single-emitter science have opened new ways to produce intensity-squeezed states, i.e., states with lower intensity fluctuations compared to coherent states. A notable 2.2 dB intensity squeezing from single-molecule emission has been demonstrated[141]. Such approaches using single-photon emitters benefit from a high capacity for spectral multiplexing due to narrow emission linewidths at cryogenic temperatures and straightforward tuning via Stark fields. An early overview and perspective on this "photon gun" approach was given in ref. 142. Epitaxial quantum dots can already produce around a few million single-photons per second with high reliability[143], and may therefore be employed in future illumination schemes, with the aim to increase the SNR by reducing uncertainty from the number of photons used in illumination. Realizing this potential would effectively close the loop between quantum emitter development and their application as active light sources in quantum microscopy.

**Outlook**

The possibilities of practical photon-resolved methodologies are vast, yet improvements in several areas are needed to realize their full capabilities.

Time-tagging array detectors are key for high-throughput higher-order correlation measurements and spatio-temporal correlation imaging. However, current SPAD arrays still lag CCD or CMOS cameras in pixel number, cost, and technological maturation[3]. For faithful resolution of correlations with n>2, the detection efficiency of SPAD arrays must be substantially improved, as the signal of the correlation function scales with the efficiency raised to the n-th power[144]. This will require pixel fill factors approaching unity. However, denser pixel spacing increases the crosstalk probability, quickly degrading higher-order correlations, and demanding innovative solutions. While near-unity detection efficiency[6] is already achieved for single-pixel SNSPDs, scalable arrays are still at the research stage and prohibitively expensive[6,79,145]. Once matured, SNSPD arrays are expected to deliver superior quantum efficiency across the UV to the infrared range, including the vibrational fingerprint region, picosecond time-resolution, near-absent dark counts, and improved ease of use in free-space mode[146]. These capabilities will realize, at full performance, many of the photon-resolved methods now being prototyped with SPAD arrays.

Data transfer and analysis methods need to keep pace with this performance, overcoming currently limited bandwidths. Bright sources of entangled photons already produce above $10^9$ pairs/second[16,147]. Fully time-tagged recording of these in imaging easily generates gigabytes per second, necessitating either efficient storage, or on-the-fly processing to immediately compress the data into specific correlations. An exciting development in this context is the adaptation of statistical learning methodologies to accelerate quantum optics experiments[148], which is just beginning to percolate into microscopy[149] and spectroscopy[150]. It remains to be seen whether quantum information science e.g., use of quantum Fisher information or quantum machine learning will have an impact in this area.

The specific requirements for photon pair sources extend beyond entanglement fidelity, state purity, and photon indistinguishability, and include higher photon flux, broaden wavelength tunability, and access to higher-N states. Generating entangled photons in the visible is of particular importance to integrate with most existing characterization methods and materials. However, visible photon sources are much less mature than their (near-)infrared equivalents with long-established use cases in optical quantum communication. Emerging methods show potential, but boosting brightness, spectral range, and robustness may require interdisciplinary breakthroughs merging quantum optics with materials science[147,151–154]. Here, a notable example is the generation of broad bandwidth entangled pairs with simultaneously reduced phase matching requirements using liquid crystals[17]. Deterministic sources based on epitaxial quantum dots, which utilize an exciton cascade for entangled photon generation, are also potential candidates with emission rates of about $10^7$ photons per second[155,156]. More complex quantum states, such as pathway entangled multi-photon states e.g., N00N states with N > 2 photons, have only been successfully demonstrated in a limited number of experiments[127,128,157–163], and foundational innovation in sources are needed first for imaging e.g., with super-sensitivity[129].

The field must also improve how it quantifies performance and defines figures of merit for new methods. A recurring concern in quantum imaging is whether the practical advantages of quantum techniques surpass classical methods, for instance when increasing the classical illumination intensity could offer higher SNR. In this context, it is important to quantify and articulate the concrete advantages of individual methods, while also drawing a clear distinction between genuinely quantum techniques and classical photon-correlation approaches more rigorously. For instance, photon-number resolved spectroscopy and microscopy proves unique utility in specific use cases —such as the unambiguous identification of optical nonlinearities in systems with a discrete number of excitations. In super-resolution microscopy, adding anti-bunching as contrast to already used intensity correlations, e.g., in SOFI, expands the utility to existing methods without promising to replace them. Such multi-dimensional quantum microscopy using anti-bunching may be used in the study of quantum emitters at low temperatures, where classical intensity fluctuations are suppressed.

The pros and cons of methods using quantum illumination likewise need to be spelled out in detail. For example, while imaging with undetected photons may support the use of less expensive and more sensitive detectors in the near- instead of mid-infrared, the spatial resolution of the captured image is still constrained by the diffraction limit of longer wavelength mid-IR light[164,165]. In cases where higher power classical illumination is limited—e.g., due to ohmic heating in plasmonic nanocavities—quantum methods surpassing the shot-noise limit can offer a unique solution. In all cases, the classical/quantum SNR comparison is essential; not only from a practical viewpoint, but also to avoid undue generalizations about the advantage of going "quantum".

Finally, we point to the need for an improved combined theoretical[166]/experimental understanding of the interaction between quantum light and complex matter. Several theoretical groups have addressed, for example, the problem of Fock state absorption[167] or quantum control over excitations via entanglement-controlled excitation pathway interference[168,169]. However, on the experimental side, nonlinear spectroscopic experiments are still outstanding, largely due to low signals from low brightness sources. Even precise and consistent measurements of the entangled-photon absorption cross-sections presents a substantial ongoing community effort in need of tight-knit theory/experiment collaborations. A critical task is the formulation of accessible frameworks describing the matter/quantum-light interaction, which challenges the common classical or semiclassical intuition of light as oscillating fields.

In conclusion, photon-resolved microscopy and spectroscopy represent a transformative frontier, pushing optical characterization methods to the limits of obtainable information at the level of individual photons. When combined with statistical learning capable of extracting hidden correlations from noisy, information-rich photon-resolved data, these methods democratize decades of quantum optics advances in bespoke methodologies as varied as their use cases -- many of which have yet to be discovered by the informed reader.

**Figures:**

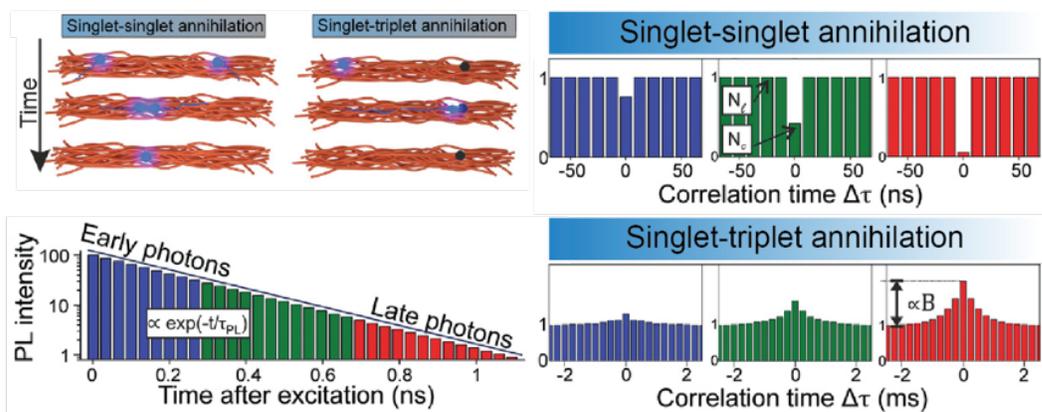

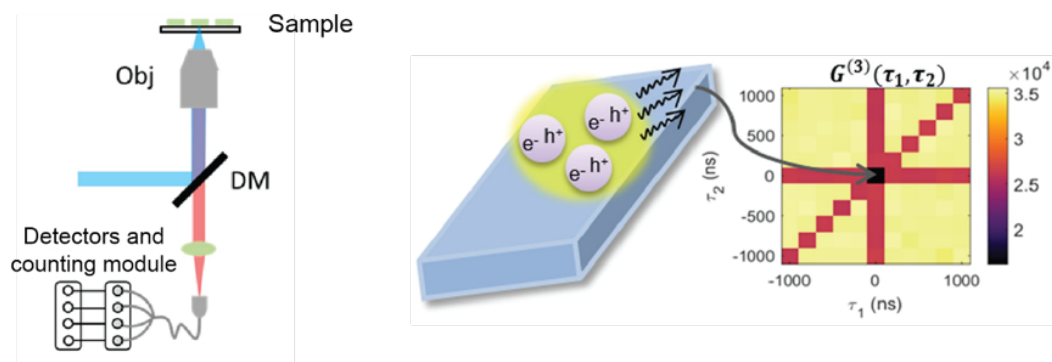

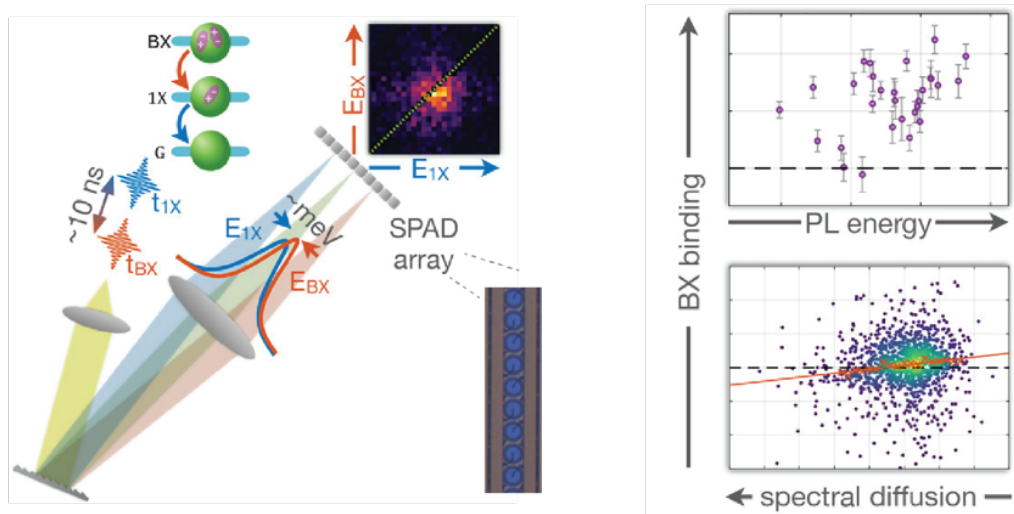

Fig. 1: **Photon-correlation for nanomaterials characterization.** (a) Time-resolved photoluminescence (PL) is obtained by measuring the time delay between the pulsed laser's electrical trigger and the subsequent PL emission, typically revealing an exponential decay that corresponds to the PL lifetime. Photons can be categorized based on their arrival times, enabling the calculation of $g^{(2)}(\tau)$ for selected photon groups, represented in blue, green, and red. This approach, known as time-resolved $g^{(2)}(\tau)$ measurements. By analyzing changes in features of $g^{(2)}(\tau)$ across nanosecond to millisecond timescales, exciton–exciton annihilation dynamics can be probed—for example, in self-assembled polymer nanoparticles[69]. (b) Recent higher-order photon-correlations can selectively probe three- or four-body excited states in semiconductor nanomaterials via the left setup. For example, third-order photon-correlations are used to study three-photon events in nanoplates[75]. The Obj refers to the objective lens, and the DM stands for the dichroic mirror. (c) Spectrally resolved photon-correlations using SPAD arrays can also measure photon-correlations between photons from different excitonic states to provide information about many-body exciton binding energies, e.g. of the biexciton in semiconductor quantum dots, and spectral diffusion dynamics[15].

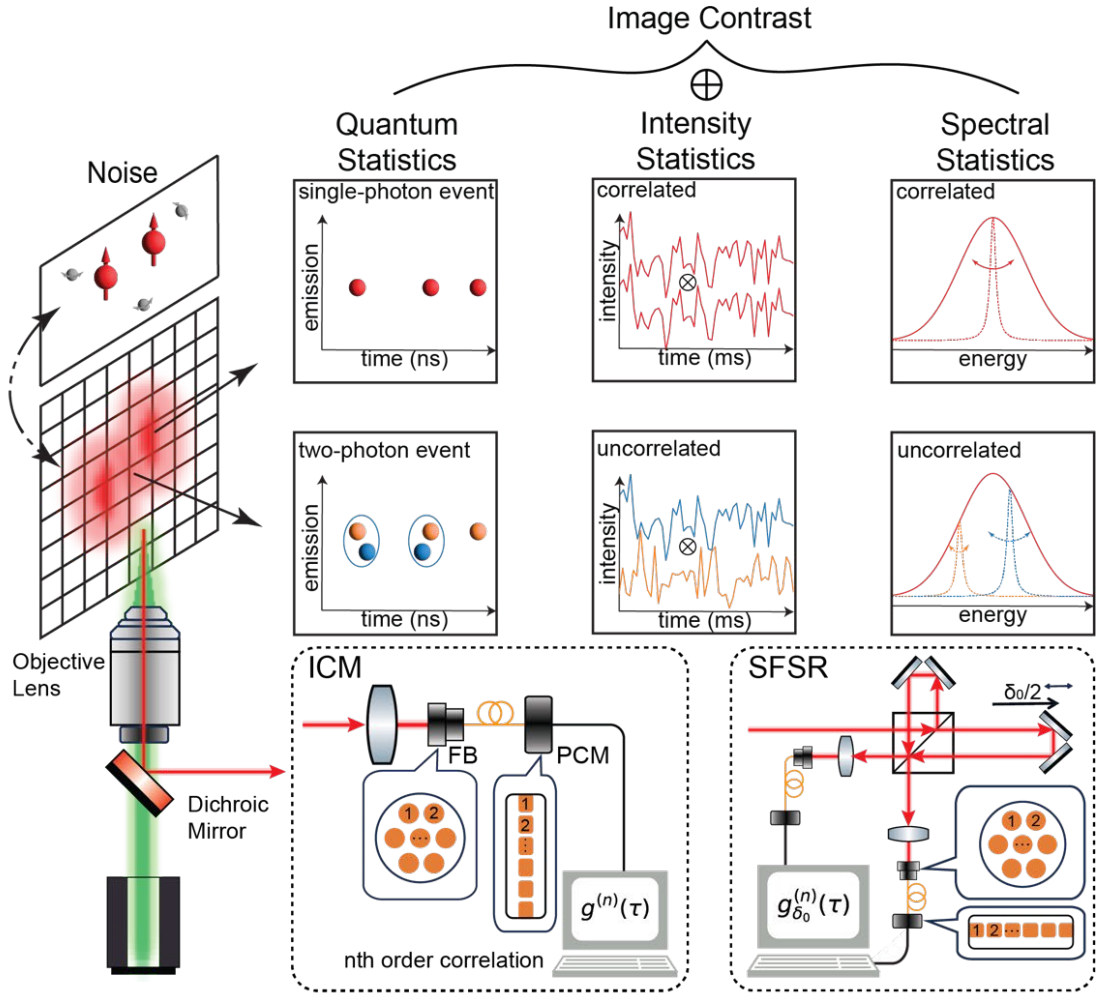

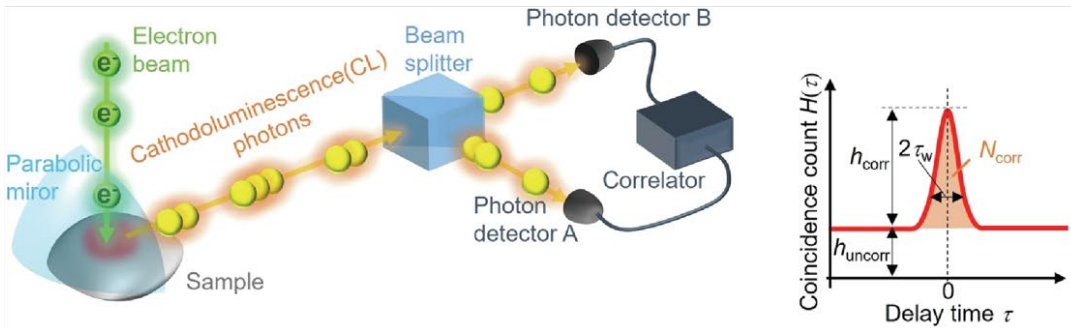

Fig. 2: **Photon-correlation in super-resolution microscopy.** (a) Principle of photon correlation microscopies. A photon-resolving detector array images two emitters undergoing independent fluctuations in their emission intensity and frequency due to nanoscale noise. The spatio-temporal correlation functions encode the overlap of emitter point spread functions (PSF); pixels receiving from multiple emitters show lower correlations than pixels receiving primarily from one emitter. Due to the nonlinearity of photon correlations, the effective PSF is narrowed, which can be leveraged for super-resolution imaging. Methods using image contrast based on quantum

and intensity temporal statistics are referred to as ICM, while that utilizing spectral statistics was introduced as SFSR. (b) Schematic illustration of photon statistics measurement in cathodoluminescence (CL)[96]. CL photons generated by free-electron excitation are directed into a Hanbury Brown–Twiss (HBT) interferometer, producing a coincidence histogram $H(\tau)$, often can be normalized as the second-order correlation function $g^{(2)}(\tau)$. A typical CL $H(\tau)$ exhibits photon bunching, which arises from synchronized emission events involving multiple emitters and reflects the underlying excitation mechanism of CL. The measured signal consists of an uncorrelated background component $h_{uncorr}$ and a correlated component $h_{corr}$, characterized by the number of correlated counts $N_{corr}$ and correlation time width $\tau_w$.

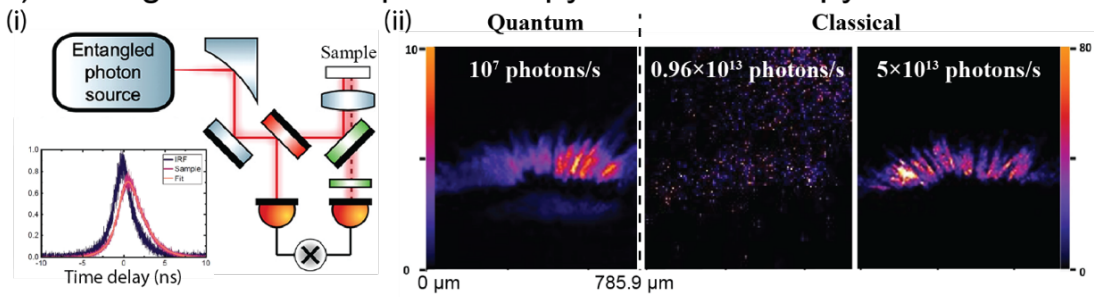

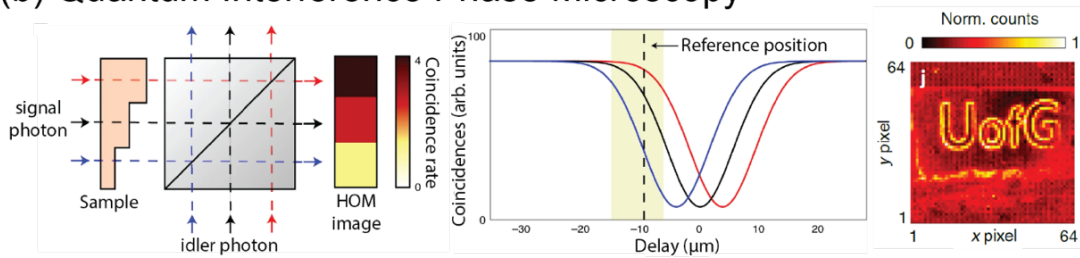

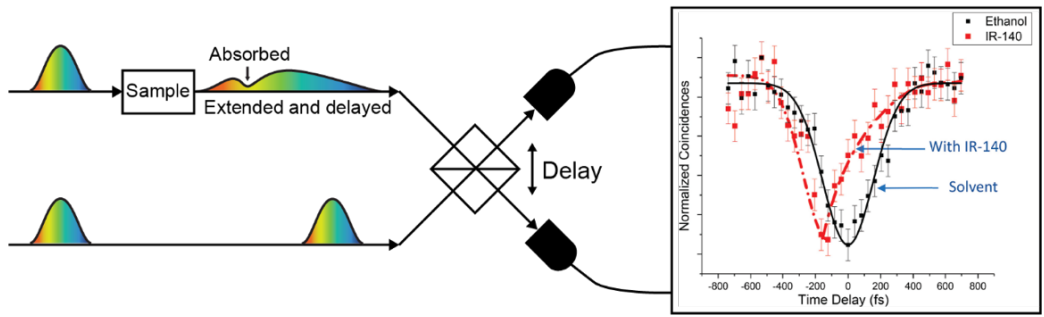

Fig. 3: **Nanosystem characterization under quantum light illumination.** (a)(i) Entangled photon spectroscopy measuring sample emission lifetimes[99]. One photon from an entangled pair is detected as herald, while its partner is focused onto a molecular sample using an objective lens. The backward-emitted fluorescence (dashed red line) is directed through a series of long-pass filters to a second detector. The sample's lifetime can be determined from the slope in $g^{(2)}(\tau)$ compiled between the two detectors. (ii) Spatio-temporal correlations between SPDC photons boost the two-photon absorption, leveraged in exceptionally low-fluence two-photon microscopy[104]. The fluorescence images generated using entangled photon pairs (labeled "quantum") demonstrate better image quality compared to those obtained with classical laser excitation (labeled "classical") despite lower excitation flux. (b) HOM-phase-contrast microscopy[110]. In HOM, two indistinguishable photons (signal, idler) coalesce at a beam splitter. The signal photon traverses a transparent sample of varying thickness profile, leading to a spatially varied HOM dip position, from which the sample topology can be reconstructed. The pixel pitch is 150 μm. (c) An emerging method in quantum materials spectroscopy sends one photon of an entangled pair through a sample medium before two-photon HOM interference. Sample-induced absorption and dispersion distort the HOM dip's shape and position. Shown on the right is experimental data from IR-140 dye molecules[113].

## (a) Quantum Imaging and Microscopy Schemes

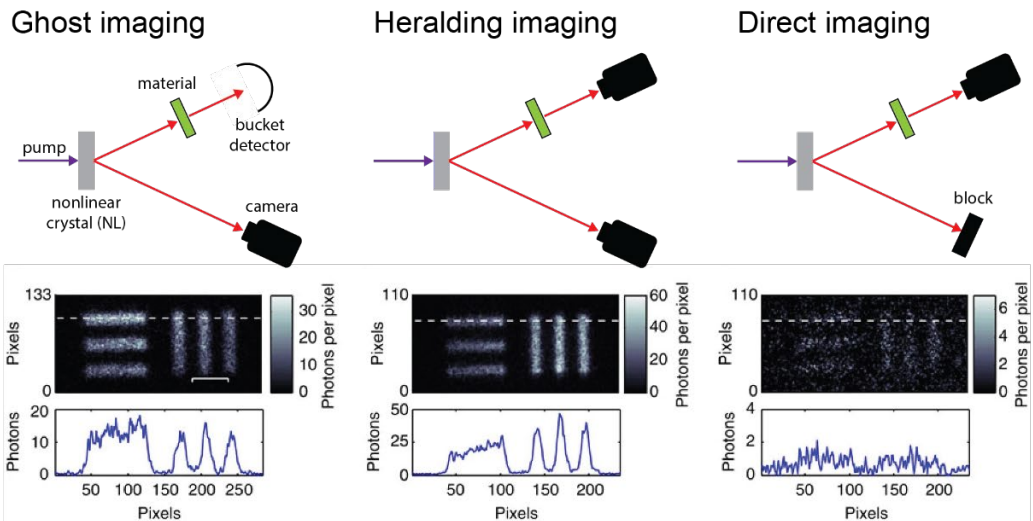

### Imaging with undetected photons

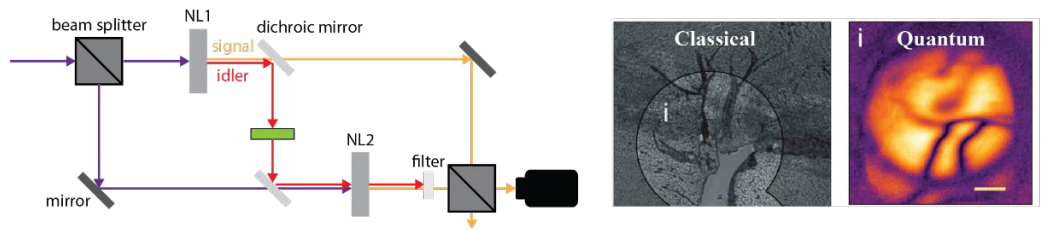

## (b) N-Photon Phase Microscopy

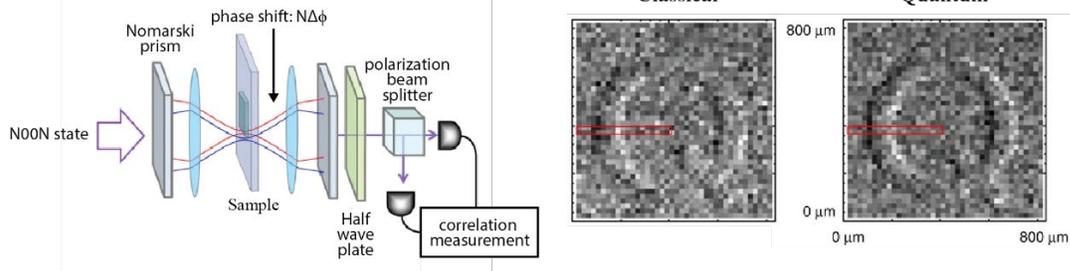

## (c) Hyperentangled Photon Microscopy

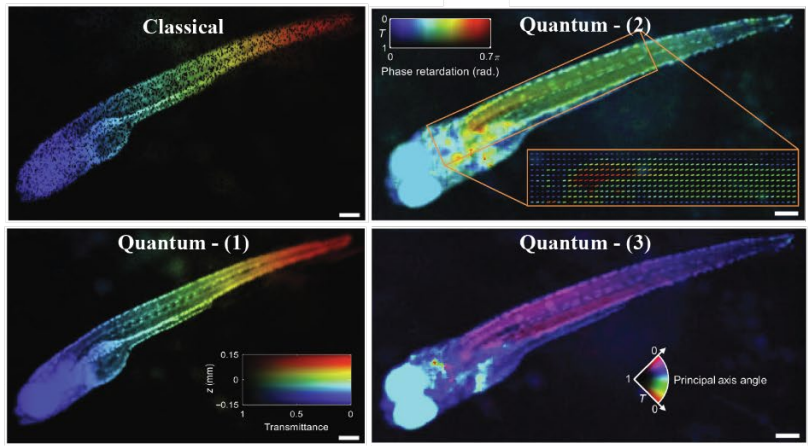

## (d) Squeezed Raman

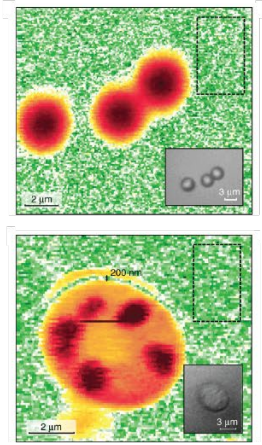

Fig. 4: **Quantum light illumination used in microscopy**. (a) Common experimental schemes used in quantum light illumination microscopy: Ghost imaging (GI) configuration, heralded imaging (HI) configuration, and direct imaging (DI) configuration as a classical benchmark[116]. Clear, high-contrast images are observed in both the GI and HI configurations, while the random nature of the detection mechanism in the DI configuration results in a low-contrast image. Scale bar: 650 µm. The right panel shows the IUP configuration: mid-IR microscopy of a mouse heart using undetected photons for absorption imaging[126]. Scale bar: 200 µm. (b) The setup and sample image under N00N state and classical light illumination in differential interference contrast microscopy[130]. (c) Hyperentangled photon pairs can form images with different degrees of freedom[136]. The classical image utilizes transmittance for contrast, whereas the three quantum images in HI configurations leverage (1) transmittance, (2) phase retardation, and (3) principal refractive index axis angle with sub-shot-noise algorithm for contrast. Scale bars, 200 µm. (d) The right panel shows images of polystyrene beads (top) at a Raman shift of 3,055 cm$^{-1}$ and a live yeast cell (bottom) at a Raman shift of 2,850 cm$^{-1}$ with squeezed light[140].

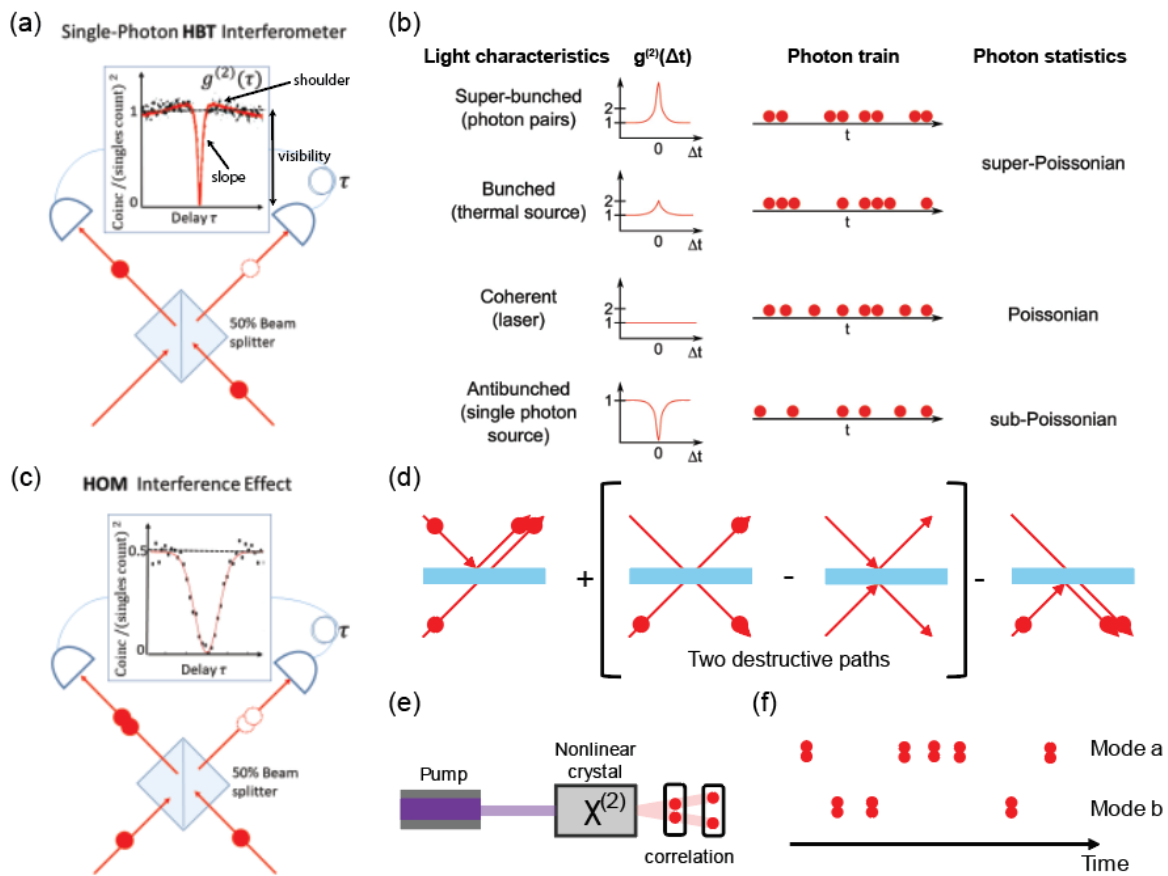

Figure Box 1. (a) Sketch of the HBT experiment[170]. In the case of a single-photon source, a photon impinging on a 50/50 beam splitter has an equal probability of being reflected or transmitted, resulting in antibunching at g$^{(2)}$(0) (b) Different light sources show distinct photon

trains, indicating various photon statistics[95]. (c) Schematic of the Hong-Ou-Mandel setup[170]. (d) When two indistinguishable photons arrive simultaneously at the input ports of a 50/50 beam splitter, quantum interference causes two of the four possible outcomes to cancel, resulting in both photons always exiting through the same output port. This phenomenon, known as the Hong–Ou–Mandel (HOM) effect, leads to the suppression of coincidence detection and is a hallmark of two-photon quantum interference. (e) In spontaneous parametric down-conversion (SPDC), a pump laser interacts with a nonlinear crystal under phase-matching conditions to emit photon pairs that are correlated in time and frequency. (f) N = 2 N00N states are quantum superpositions of two photons -- either both in one optical mode or both in the other, but never split between the two.

## BOXES

BOX 1 Quantum Optics Terminology

1. Photon statistics, temporal photon correlations, and photon number-resolved measurements.
   Photon correlation measurements analyze the temporal relationships between individual photon detection events, often after optical manipulation such as passage through color-, or polarization filters, or interferometers. By "time-tagging" the arrival of photons and computing correlations, these measurements yield a concise observable known as the correlation function e.g. $g^{(2)}(\tau)$ for the second-order correlation function, where $\tau$ represents the lag time between photon pairs. Photon correlations are the default way of extracting the underlying temporal photon statistics of light.
2. Hanbury Brown-Twiss (HBT) interferometer and $g^{(2)}(\tau)$
   The HBT setup is a widely used configuration for measuring second-order photon correlation functions[171]. A 50/50 beam splitter directs the photon stream onto one single-photon detector at each output port, respectively. In the language of quantum optics, the correlation function for a single field mode is written in terms of photon creation and annihilation operators:

$$g^{(2)}(\tau) = \frac{\langle \hat{a}^\dagger(t)\hat{a}^\dagger(t+\tau)\hat{a}(t+\tau)\hat{a}(t)\rangle}{\langle \hat{a}^\dagger(t)\hat{a}(t)\rangle\langle \hat{a}^\dagger(t+\tau)\hat{a}(t+\tau)\rangle} \quad (1)$$

More generally, it can also be written using $\hat{E}^{(+)}(t)$ and $\hat{E}^{(-)}(t)$, the positive and negative frequency components of the electric field operator, respectively. The ordering of operators must be considered, requiring the use of normal ordering notation.

$$g^{(2)}(\tau) = \frac{\langle :I(t)I(t+\tau): \rangle}{\langle I(t)\rangle\langle I(t+\tau)\rangle} = \frac{\langle \hat{E}^{(-)}(t)\hat{E}^{(-)}(t+\tau)\hat{E}^{(+)}(t+\tau)\hat{E}^{(+)}(t)\rangle}{\langle \hat{E}^{(-)}(t)\hat{E}^{(+)}(t)\rangle\langle \hat{E}^{(-)}(t+\tau)\hat{E}^{(+)}(t+\tau)\rangle} \quad (2)$$

In this quantum interpretation, $g^{(2)}(\tau)$ is defined in terms of coincidences between photon counting events, the number of which per time also relate to the classical time-dependent intensity, $I(t)$. In the classical context, $g^{(2)}(\tau)$ can be defined in terms of the intensity fluctuations of the incident light in a classical interpretation:

$$g^{(2)}(\tau) = \frac{\langle I(t)I(t+\tau)\rangle}{\langle I(t)\rangle\langle I(t+\tau)\rangle} \quad (3)$$

, where τ is the time delay, $I(t)$ is the light intensity at time t. Different archetypes of $g^{(2)}(\tau)$ are observed depending on the source characteristics[172]. Perhaps counterintuitively, a perfectly coherent source e.g., a laser, exhibits a random (Poissonian) distribution with $g^{(2)}(\tau) = 1$. $g^{(2)}(\tau) > 1$, indicates photon "bunching", and photon pairs appear more frequently than in the random distribution. $g^{(2)}(\tau) < 1$, it is termed "antibunching," signifying that photon pairs appear less frequently than in the random distribution, e.g., observed in spontaneous emission of two-level systems – "quantum emitters".

3. Hong-Ou-Mandel (HOM) effect

   When two indistinguishable photons enter a lossless beam splitter from two separate input ports, there are four possible outcomes for the output: both photons are transmitted, both are reflected, one is transmitted (reflected) while the other is reflected (transmitted). Due to energy conservation, there is an overall π-phase shift in the last two cases (some articles describe this using two indistinguishable alternatives[144,173,174]), leading to destructive interference of the probability amplitude of photons in different output modes. This phenomenon is called Hong-Ou-Mandel (HOM) interference, also known as two-photon interference or fourth-order (field) interference[175]. By placing two detectors at the output ports and performing a coincidence measurement at various time delays between the two photons before they interfere, a characteristic dip is observed at zero-time delay, known as the HOM dip. It is usually used to measure the indistinguishability of photon pairs and generate other quantum states of light in metrology.

4. Spontaneous parametric down-conversion (SPDC)

   Spontaneous parametric down-conversion is widely used to generate entangled photon pairs[8]. In the process, a high-energy photon, or pump photon, is converted into a pair of lower-energy photons known as daughter photons: a signal photon and an idler photon. When the daughter photons have the same energy, they are termed degenerate photon pairs; otherwise, they are nondegenerate.

5. N00N states

   A N00N state, also known as a NOON state, is a quantum-mechanical many-body entangled state. It can be written as:

   $$\frac{|N\rangle_a|0\rangle_b + |0\rangle_a|N\rangle_b}{\sqrt{2}}$$

   Such state represents a superposition of N particles in mode "a" with 0 particles in mode "b", and 0 particles in mode "a" with N particles in mode "b", which can exhibit path dependence, polarization dependence, and other variations depending on the chosen mode. The most common N00N state is a path-dependent N = 2 N00N state. It can be generated by passing entangled photon pairs through a HOM interferometer[129]. Due to the HOM effect, two photons will bunch together and travel to the same output port. Higher N N00N states can be generated by mixing coherent light states with entangled photon pairs, leading to polarization dependent higher N00N states[163].

6. Squeezed states

   A squeezed state is a type of quantum state of light where the quantum uncertainty (noise) in one property, such as intensity or phase, is reduced below the standard quantum limit at the expense of increased uncertainty in the conjugate property[172]. This process, known as "squeezing," is usually used to enhance precision measurements.

BOX 2 advanced detectors and parameters

1. Single-Photon Avalanche Diodes (SPADs)

   Also known as Geiger-mode avalanche photodiodes (APDs), these detectors operate with a bias voltage above their avalanche breakdown threshold[2]. A Single electron-hole pair after absorption of a single photon can trigger an impact-ionization process, resulting in an avalanche current that marks the detection of a single photon. The timing jitter is on the order of tens to hundreds of picoseconds. Shortwave-infrared InGaAs APDs are still lagging in quality, often showing higher (>150 cps) dark counts compared to silicon analogs (30-60cps).

2. SPAD arrays

   Unlike conventional multi-pixel detectors e.g., EMCCDs that only record intensity traces at millisecond resolution, SPAD arrays provide time-tagged detection of individual photons using a multitude of APDs on the same chip. SPAD arrays allow high-speed imaging and enable pixel-wise correlation analysis across the array[3]. Despite their advantages, current SPAD arrays are limited by factors such as low spatial fill factor (pixel density), number of pixels, substantial dark counts (>1,000 for some individual pixels), and up to 2% crosstalk probability between neighboring pixels.

3. Superconducting Nanowire Single-Photon Detector (SNSPD)

   A typical SNSPD consists of a superconducting film patterned into a wire with nanoscale dimensions. These detectors are generally fabricated from a film approximately 5 nm thick, shaped into a meandering wire about 100 nm wide to maximize surface area[176]. Compared to SPADs, SNSPDs exhibit superior performance in the infrared range[7]. They also offer higher photon detection efficiency (>90%), a lower dark count rate (<1 cps), and minimal timing jitter (tens of ps)[6]. SNSPDs require liquid helium cooling and fiber-coupling of the optical input, thereby introducing experimental complexity. SNSPDs can also be integrated into array-type detectors, although they are still in the early stages of development. One significant limiting factor in achieving large-scale arrays is the readout architecture[177,178], which must be highly efficient to avoid raising the device temperature.

   Key metrics in single-photon detection

   a. Photon detection efficiency

      Quantifies the conversion efficiency of photons into detectable photoelectrons. It represents the overall efficiency of the detection process.

b. Dark count rate
   The average rate of registered counts in the absence of any incident light, known as the dark count rate, determines the minimum count rate at which the signal is predominantly caused by real photons.
c. Timing jitter
   Refers to the temporal uncertainty or variation in the arrival time of a detected photon or voltage pulse relative to an expected or ideal arrival time. This deviation arises due to factors such as noise, variations in the detector response time, or fluctuations in the electronic circuitry, which can directly affect the temporal resolution of photodetectors.
d. Figure of merit
   Introduced as a parameter to assess the quality of photodetectors, it is defined as the ratio of the photon detection efficiency to the product of the dark count rate and timing jitter.
e. Dead time
   The minimum time interval required between two consecutive counts for them to be recorded as separate events.
f. Maximum counting rate
   Refers to the speed at which a detector can respond to incoming photons. It is defined as the reciprocal of the dead time.
g. Fill factor
   Refers to the ratio of the active area available for photon detection to the total area of the detector, with any metal layers overlapping the active area subtracted, that is, the ratio of sensitive and insensitive area. It quantifies the effective utilization of the detector's surface for detecting photons.
h. Crosstalk
   A phenomenon that occurs in array-type photodetectors, where the detection of a photon by one pixel can influence nearby pixels, resulting in additional pulses.

## Acknowledgements

This work was financially supported by the College of Chemistry at the University of California, Berkeley and the U.S. Department of Energy (DE-AC02-05CH11231).